\documentclass[12pt]{article}

\usepackage{newtxtext,newtxmath}

\usepackage{graphicx}

\usepackage[letterpaper,margin=1in]{geometry}

\renewenvironment{abstract}
	{\quotation}
	{\endquotation}

\date{}

\makeatletter
\renewcommand{\fnum@figure}{\textbf{Figure \thefigure}}
\renewcommand{\fnum@table}{\textbf{Table \thetable}}
\makeatother

\usepackage{scicite}

\usepackage{url}

\usepackage{aasmacros}

\def\scititle{
	The discovery and scientific potential of fast radio bursts
}
\title{\bfseries \boldmath \scititle}

\author{
	Matthew~Bailes$^{1\ast\dagger}$\\
	\small$^{1}$Centre for Astrophysics and Supercomputing, \\
	\small Swinburne University of Technology,\\ 
	\small Hawthorn, Vic, 3122, Australia.\\
	\small$^\ast$Corresponding author. Email: mbailes@swin.edu.au\\ 
}

\begin{document} 

\maketitle 

\begin{abstract} \bfseries \boldmath
Fast radio bursts (FRBs) are millisecond-timescale bursts of coherent radio
emission that are luminous enough to be detectable at cosmological distances. 
In this review I describe the discovery of FRBs, subsequent advances in our understanding of them, and future prospects.
Thousands of potentially observable FRBs reach Earth every day; they probably originate from highly magnetic and/or rapidly rotating neutron stars in the distant Universe.
Some FRBs repeat, with this
sub-class often occurring in highly magnetic environments. Two repeaters exhibit cyclic activity windows, consistent with an orbital period. One nearby FRB was from a Galactic magnetar during an X-ray outburst. The host galaxies of some FRBs have been located, providing information about the host environments and the total baryonic content of the Universe.

\end{abstract}

\noindent
Fast radio bursts (FRBs) are millisecond flashes of radio waves from distant astronomical sources.
The first FRB \cite{2007Sci...318..777L}, later known as the Lorimer Burst,
was discovered in 2007, some forty years after the discovery of radio pulsars\cite{1968Natur.217..709H}.
As was the case with pulsars, its discovery was 
completely serendipitous. The burst swept across the 288 MHz passband of the receiver over about 1/3 of a second, consistent with radio waves  from a distant extraterrestrial source, dispersed by passage through an ionised plasma (Figure \ref{fig:DM} D-E). It struck the Parkes 64-m radio telescope in Australia just after 5:50 AM on August 25, 2001 local time, and was automatically archived
onto magnetic tape as part of a radio pulsar survey. It remained undiscovered for over five years. This review examines
its serendipitous discovery, the subsequent demonstration that it was part of a new  class of object, and their  cosmological implications.


\subsection*{Discovery of the Lorimer Burst}

The discovery of FRBs became possible due to development of
high time resolution radio instrumentation and software tools, mainly used for pulsar surveys.
Radio pulsars are rapidly-rotating highly-magnetised neutron stars that emit beams of emission that appear to pulse as they rotate \cite{1968Natur.217..709H,
1969ApJ...157..869G,
1972ARA&A..10..427R,2012hpa..book.....L}, like a light house.
By 2007 over half of known pulsars had been discovered using the
Parkes 64-m telescope (Figure \ref{fig:Montage}A), due to its
relatively radio interference-free environment, and access
to the Southern hemisphere sky (where the majority of pulsars reside)
\cite{2001MNRAS.328...17M,2002MNRAS.335..275M,2003MNRAS.342.1299K,2001ApJ...553..367C}. 

The study of radio pulsars has been advanced by discoveries of unusual objects, 
usually in large-scale surveys\cite{1968Natur.217..709H,1975ApJ...195L..51H,1982Natur.300..615B}. A new class of pulsar-like objects, the rotating radio transients (RRATs) were discovered in 2006 using the Parkes Multibeam Receiver\cite{1996PASA...13..243S}, a 
13-pixel radio camera which was used in multiple pulsar surveys\cite{2001MNRAS.328...17M,2002MNRAS.335..275M,2001MNRAS.326..358E, 2006ApJ...649..235M}.
RRATs\cite{2006Natur.439..817M} were initially interpreted as a new type of pulsar that only rarely emitted
pulses, albeit in phase with a neutron star's rotation period.
This was unlike the radio pulsars, that usually emitted regular 
pulsations with each rotation. 
In 2007 searches were underway to find more example RRATs\cite{2006ApJ...645L.149W,2010MNRAS.401.1057K,2010MNRAS.402..855B}.

\begin{figure}
	\centering

\includegraphics[width=0.9\textwidth]{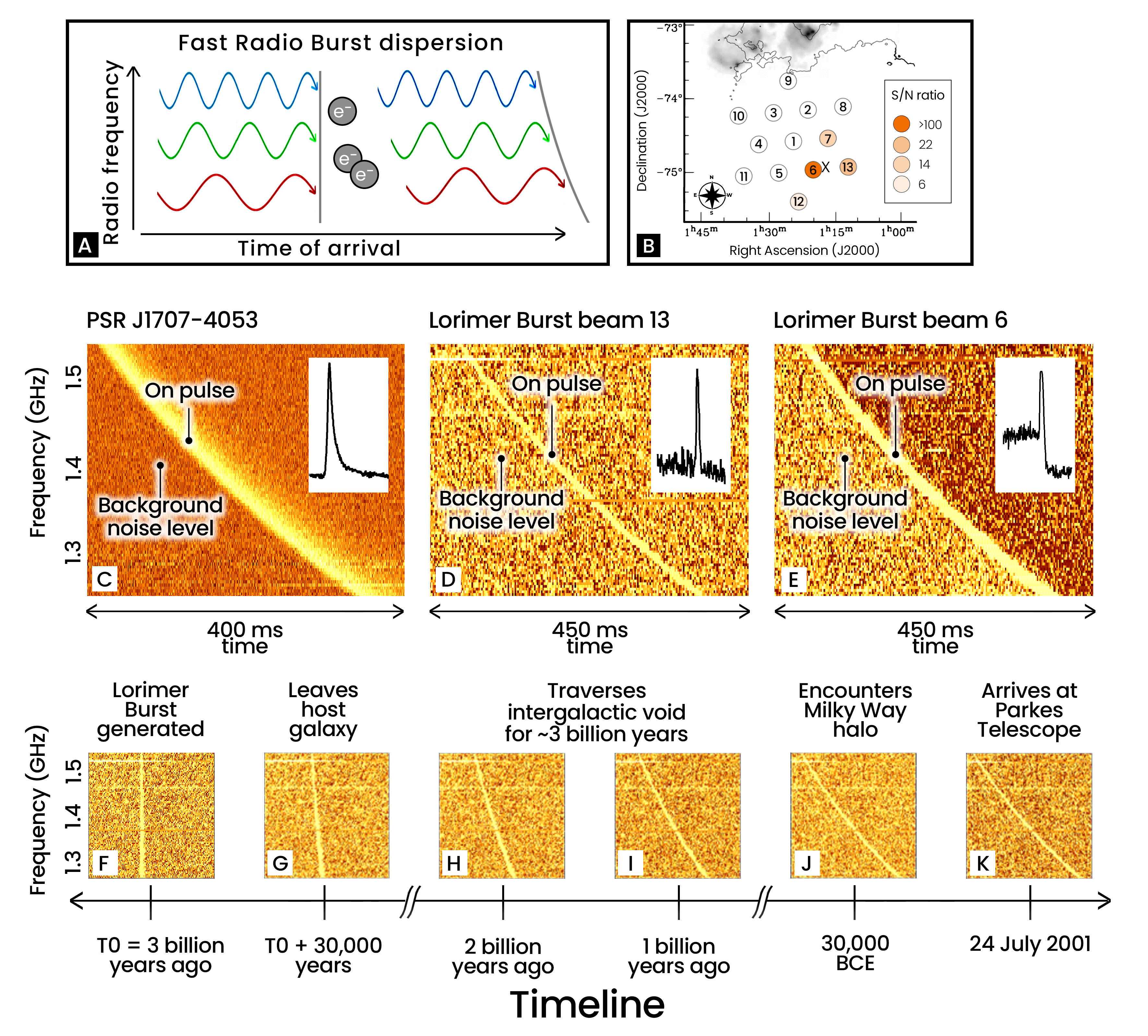} 

	\caption{\textbf{Fast radio burst dispersion and the location of the Lorimer Burst.}
		(A) Conceptual illustration of how dispersion delays the time of arrival at Earth. As radio waves encounter free electrons, they become delayed in a
		radio frequency-dependent manner. The more energetic (higher frequency) 
		radio waves (blue) experience less delay than
		the lower energy waves (red).
		This leads to a characteristic sweep, observed in
		fast radio bursts and pulsars. (B) The pointing of the Parkes 13-beam receiver just south of the edge of the Small Magellanic Cloud at the time of the observation of the Lorimer Burst\cite{2007Sci...318..777L}. The burst saturated beam 6, was well above the detection threshold in beams 7 and 13, and was weakly detected in beam 12\cite{2019MNRAS.482.1966R}. The cross indicates the inferred burst position\cite{2007Sci...318..777L}.
		(C) Observed dispersion sweep of
		the pulsar PSR J1707--4053 with $DM=360$ pc cm$^{-3}$ and its de-dispersed
		pulse profile (inset) with the MeerKAT telescope\cite{2020PASA...37...28B}. (D-E) Dispersion
		sweep (and inset integrated pulse profile) of the Lorimer burst, at $DM=375$ pc cm$^{-3}$ 
		in sidelobe beam 13 (D)
		and beam 6 (E). The dip in flux after the burst in beam 6 is an instrumental artefact due to saturation. (F-K) 
		Inferred evolution of the Lorimer Burst's dispersion over cosmic time. 
		}
	\label{fig:DM} 
\end{figure}

Like pulsars, RRATs exhibit a radio frequency-dependent delay, which appears as a sweep
on the radio receiver. This arises because in 
the ionised interstellar medium radio waves travel slightly slower than the speed of light in a vacuum ($c$). The 
speed is determined by the radio frequency $\nu$ and the density of free electrons (Fig \ref{fig:DM}A). 
By the time a broadband radio pulse arrives at Earth,
this frequency-dependent delay leads to a well defined sweep, in which the higher radio
frequencies arrive before the lower ones (shown in Fig \ref{fig:DM}C for PSR J1707--4053). 
A radio pulse's dispersion measure $DM \equiv \int_0^L n_{\rm e} dL$ is the integrated column density of free electrons along the line of sight in pc cm$^{-3}$, where pc is the parsec, cm is the centimetre, $n_{\rm e}$ is the local free electron density, and $L$ is the distance to the source. The difference in the arrival times $t_2-t_1$ between two radio waves
of radio frequencies $\nu_2$ and $\nu_1$ respectively in GHz is 
approximately\cite{2020arXiv200702886K}
\begin{equation}
    t_2-t_1 \approx 4.15 \left[\left(\frac{1}{{\nu_2}^2}\right)-\left(\frac{1}{{\nu_1}^2}\right)\right] \,DM \,\rm{ms}
\end{equation}
\noindent 
where the $DM$ is in the units pc cm$^{-3}$ and the radio frequency is in GHz.
The measured $DM$ of a source, combined with a free electron model, can be used as a 
proxy for distance within our  Galaxy\cite{2002astro.ph..7156C,2017ApJ...835...29Y} and to cosmological distances\cite{2003ApJ...598L..79I}. 
Although it greatly complicates radio pulsar instrumentation and analysis, pulse dispersion helps observers distinguish celestial sources from local radio interference, and is ultimately what allows FRBs to have cosmological applications.

Duncan Lorimer and undergraduate student Narkevic searched for RRATs in an
archival multibeam pulsar survey of the Small Magellanic Cloud (SMC), a nearby dwarf galaxy,
that has used the Parkes 64-m telescope.
Narkevic's initial analysis had detected a total of two putative RRATs in the whole
SMC survey. They both appeared at exactly the
same time with $DM \sim375$ pc cm$^{-3}$ in two adjacent beams of the 
13-beam receiver, suggesting that either a very bright source was 
illuminating the sidelobes of multiple beams, or an unusual
form of radio interference (Fig \ref{fig:DM}B).

I was observing at
the Parkes telescope on an unrelated project with Lorimer and
became involved in diagnosing the putative dispersed radio source.
We extracted the relevant burst data to display the so-called waterfall plots, which would show the
pulse energy as a function of both time and frequency\cite{2004PASA...21..302H,2011PASA...28....1V}.
We found the source had a
dispersion sweep similar to those of pulsars, with evidence of radio frequency-dependent multi-path interstellar scattering (that scales as $\nu^{-4}$), as often seen in observations of pulsars\cite{2004ApJ...605..759B}  (Fig \ref{fig:DM}D-E).
Further investigation showed that the burst had saturated the limited dynamic range receivers
of the beam closest to the source's
location (beam 6 in Fig \ref{fig:DM}B)  
and an algorithm designed to remove interference 
had replaced the burst with synthetic data. If not for the multibeam receiver's sensitivity to strong signals in multiple beams the burst would never have been found.
Disabling the interference rejection algorithm and reprocessing the archived data showed that the burst was 
over 100 times the survey's detection threshold, with an estimated flux density of 30 jansky (Jy) (Figure \ref{fig:DM} D-E). 
The burst was bright enough
to be visible in four of the thirteen beams
\cite{2016PASA...33...45P,2019MNRAS.482.1966R}.

\begin{figure}
	\centering
	\includegraphics[width=0.8\textwidth]{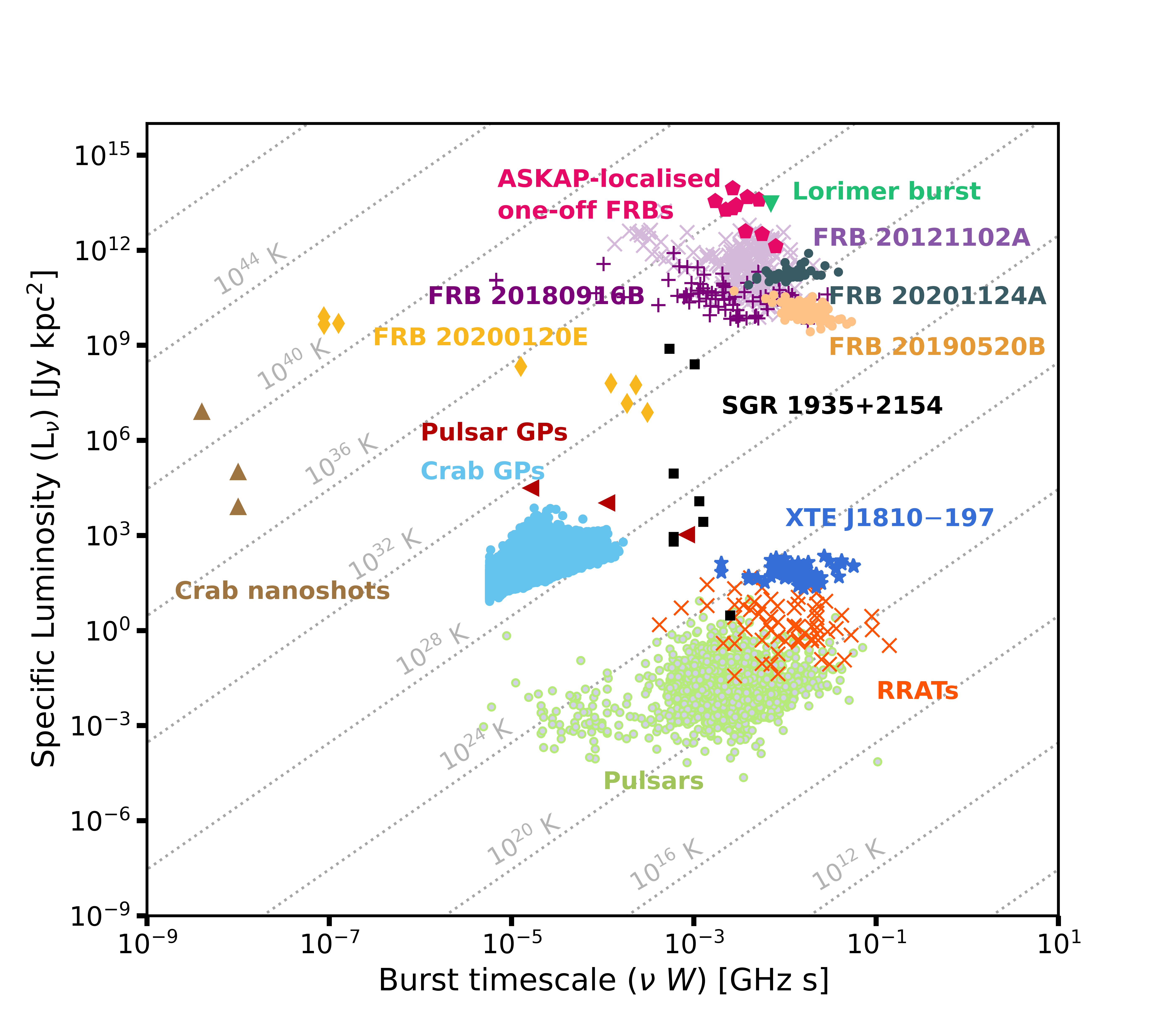} 

	\caption{\textbf{The luminosity as a function of burst timescale of
	short-duration coherent radio emitters.}
	  The Lorimer Burst (green triangle) is a   trillion times more energetic than the radio pulsar (light green circles) and
	  RRAT (orange crosses) populations of our own Galaxy. Also shown
	  are other one-off FRBs (magenta pentagons), repeating bursts from FRB 20121102A (light purple crosses) and 
	  FRB 20180916B (dark purple plus symbols),
	  the globular cluster repeating 
	  FRB in M81 FRB 20200120E (yellow diamonds), and radio bursts from
	  the galactic magnetars SGR 1935+2154 (black squares) and XTE J1810--197 (blue stars).
	  Nanosecond duration bursts from the Crab pulsar (brown triangles) possess similar brightness
	  temperatures (equivalent black body temperatures) to the most energetic FRBs.
	  The sloped dotted grey lines indicate the brightness temperature, which is proportional to the specific intensity
	  of a source. Luminosities and timescales are
        sourced from the FRB review paper\cite{2021Univ....7..453C} and references\cite{2007Sci...318..777L,2018Natur.562..386S}.
		}
	\label{fig:Phase} 
\end{figure}

The estimated source distance placed it well beyond
the SMC (the target of the survey); it
appeared to be at cosmological distance\cite{2007Sci...318..777L}.
In the voids between galaxies, free electron densities\cite{2003ApJ...598L..79I} are about 1 m$^{-3}$. 
This raises the source distance by about 1~Mpc for each pc cm$^{-3}$ of the $DM$.
For reference, the Milky Way is about 30 kpc across
and the SMC is at a distance of approximately 60 kpc.
Thus the $DM$ of $\sim$375 pc cm$^{-3}$ indicated a distance of $\sim$1 Gpc.
To explain the observed brightness at such a distance,
the source would have to be about a trillion times
more luminous than any known pulsar.
Alternatively the source could have been enshrouded 
 in a highly ionised plasma
in its host environment, leading to a spurious distance estimate.
There is a sub-class of high-magnetic field and/or short-period 
pulsars known to occasionally 
emit highly energetic radio pulses that might have been linked to the burst.
The Crab Pulsar, a young and highly-magnetised
neutron star, occasionally produces these 
`giant pulses', individual radio flashes
that can be $>>$ the mean energy of its average pulse.
However, even the brightest Crab Pulsar giant pulses were about a trillion
times less energetic than the putative burst (Figure \ref{fig:Phase}). 
The implied radio
energy in the 5-ms duration burst was similar to all the power
our Sun emits over a month. 

The burst was well above the survey detection
threshold. In extragalactic surveys
a homogeneous isotropic cosmological population follows a 
d log $N$/d log $S$ = --3/2 relation, where $N$ is the number of
sources above a flux density $S$. This arises because the volume
surveyed expands as the distance $D^3$, while each object's flux density follows the inverse square law
 $D^{-2}$. For every Lorimer
Burst there should be 
many (perhaps dozens) of fainter bursts present in the survey data. 
There didn't appear to be any\footnote{A later FRB
search\cite{2019MNRAS.484L.147Z} found another FRB in the same dataset, with a 
DM of 1187 pc cm$^{-3}$}.

If the source repeated, it would provide more confidence in
its celestial origin, but 40 hours of follow-up observations at Parkes saw nothing\cite{2007Sci...318..777L}. 
It appeared that the burst was either i)
a one-off hyper-luminous flash of radio waves from the distant Universe, a trillion times more luminous than known radio transient bursts, and requiring improbable survey statistics or ii) some obscure form
of radio interference. We 
decided that the pulse's well-defined sweep with frequency was sufficient to conclude it was celestial, so
submitted a paper summarising our findings\cite{2007Sci...318..777L}.
The event soon became known as the `Lorimer Burst', now also designated FRB 010724. 

\subsection*{Implications of the Lorimer Burst}

The apparent luminosity of the Lorimer Burst
was very high but there were few clues as to what
may have caused it.
There was only an upper limit ($\leq$5 ms) on
the intrinsic width of the burst, due to a combination of instrumental 
broadening and radio wave scattering.
Requiring the emission region to be causally connected ( i.e. within 5 ms travel at the speed of light) set an upper limit on its size of $\leq$ 1500 km.
Such distance scales are consistent with
the dimensions of rotating magnetic fields possessed by rapidly-rotating neutron stars, shocks from explosions emanating from relativistic objects, 
or collisions between neutron stars and
other compact objects (neutron stars and stellar-mass black holes).

Radio telescopes observe such a 
small fraction of the sky that initial estimates of 
the FRB event rate were hundreds per day as bright as the Lorimer Burst\cite{2007Sci...318..777L}
and 10,000 d$^{-1}$ fainter bursts which were still above the detection limit of large radio dishes.
The implied rate in a given cosmological volume was 
similar to that of supernovae:
about once every few decades in a normal galaxy.

If its origin could be determined, the Lorimer
Burst had potential for use as a cosmological probe. The $DM$ contains information on the 
number of free electrons along its path.
Most of the Universe's baryonic (normal atomic) mass is not in galaxies, but between them in the intergalactic medium (IGM). 
The bulk of the IGM mass is in hydrogen and helium, which 
cannot retain their
electrons as they are ionised by 
ultra-violet light. An FRB can therefore act as a free electron 
(and hence baryon)
counter between the host galaxy and Earth
(Figure \ref{fig:DM} F-K). Localised FRBs 
could potentially constrain the total mass of IGM, a quantity with controversial measurements via other methods\cite{2012ApJ...759...23S}.

\subsection*{The hunt for more FRBs}

Attempts were made to find more bursts, both
from existing
archival data and by initiating new surveys. 
However, none of the early efforts were  
successful. Many groups searched archival data, finding some new RRATs\cite{2010MNRAS.402..855B,2012MNRAS.425.2501B},
but no events similar to the Lorimer Burst.
Surveys specifically designed to find new bursts
\cite{2012ApJ...744..109S,2013ApJ...763...81L} also
initially did not detect any.
Later it was discovered that the Parkes radio telescope  
receiver was often struck by quasi-dispersed radio pulses
that were present in all 13 beams of the multibeam
receiver, some of which had dispersion similar to the Lorimer 
Burst\cite{2011ApJ...727...18B}. These were shown to be due to
a microwave oven only about 100\,m from the Parkes dish\cite{2015MNRAS.451.3933P}, raising further doubts. Was the Lorimer
Burst just a similar form of radio interference after all?

In 2012 another search of the Parkes Multibeam Pulsar
Survey found a potential Lorimer Burst-like signal \cite{2012MNRAS.425L..71K} 
with a $DM$=745 pc cm$^{-3}$ although it was uncertain
whether it originated outside the Galaxy\cite{2014MNRAS.440..353B}.
The celestial and cosmological nature of these 
dispersed radio signals
remained uncertain.

\subsection*{2013: A Cosmological Population}

In 2008 three pulsar and radio burst surveys commenced at Parkes, called the High Time Resolution Universe (HTRU) surveys
\cite{2010MNRAS.409..619K}.
These employed multibeam anti-coincidence detection methods\cite{2012MNRAS.420..271K} that proved effective
at removing terrestrial near-field interference. 
One of the HTRU surveys found 
four radio bursts\cite{2013Sci...341...53T}, the
brightest of which had a $DM$ almost 3 times that of the Lorimer burst
and followed the same $\nu^{-2}$ dispersion and $\nu^{-4}$
scattering power laws with radio frequency.
This demonstrated that
fainter and higher-dispersion (more distant) bursts existed, so the Lorimer Burst was part of a 
 cosmological population.
The term `fast radio bursts' and acronym (FRBs) was coined.
A system of nomenclature was adopted, designating each burst FRB followed by numerals indicating the date it occurred, similar to that used for 
for gamma-ray bursts. For example, 
FRB 110220 was detected on 2011 February 20 Universal Coordinated Time (UTC); it 
had a dispersion measure of 995 pc cm$^{-3}$ \cite{2013Sci...341...53T}.
Models of extragalactic dispersion indicated the sources could be at distances up to redshift $z\sim1$, 
when the Universe was half its current age.

\subsection*{2014-2017: Single-dish discoveries}

Although the four additional FRBs were reassuring, 
they had also been discovered with the Parkes 64-m telescope.
Was there something in its local environment that was mimicking dispersed pulses?
This doubt was dispelled in 2014 when a team 
using the Arecibo 305-m dish in Puerto Rico (Figure \ref{fig:Montage} B)
announced the detection of another FRB\cite{2014ApJ...790..101S}.
FRB 121102 had $DM=$ 556 pc cm$^{-3}$
and like the Parkes examples, appeared to be far beyond
the Milky Way. 
Another $DM$=790 pc cm$^{-3}$ FRB was found in archival data\cite{2014ApJ...792...19B}, then a bright FRB was identified using the 100\,m Green Bank
Telescope (GBT)\cite{2015Natur.528..523M}, which recorded  
full polarimetric information, showing the FRB had a polarisation fraction of almost 50\%. 
Changes in the position angle of the polarisation as a function of frequency
(known as Faraday rotation) were evident, indicating that the FRB source was
probably immersed in a highly magnetised region
within its host galaxy. Another 5 FRBs\cite{2016MNRAS.460L..30C} were
announced from the HTRU survey,
including evidence that the bursts might have
multiple components, and a  $DM$ as high as
1629 pc cm$^{-3}$. Other 
FRBs well above the detection threshold
were also discovered using the Parkes 64-m
telescope\cite{2015ApJ...799L...5R,2016Sci...354.1249R}.

Single dish radio telescopes have poor spatial resolution
($\sim$ sq degree), so none of these FRBs could be unambiguously
associated with a host galaxy. 
A definitive demonstration that 
FRBs are at cosmological distances
could potentially be made by using an interferometer,
which has much better spatial resolution, to localise an FRB to a host galaxy\cite{2014ApJ...797...70K}. 
Without known
distances or association at other wavelengths,
it was difficult to constrain models of emission
mechanisms, or even
be certain that FRBs were a cosmological population.

\subsection*{Early Physical Models}

A catalogue of theories for
the emission mechanism of FRBs\cite{2019PhR...821....1P}
listed more models than there were then known bursts.
If FRBs are at cosmological
distances, then their estimated radio energies are
10$^{38}$ to $10^{40}$ ergs. 
This is about the same total energy the Sun emits
in a day to a month, but all in the radio band, and all
within a few milliseconds. Although this eliminates Sun-like stars as the source of FRBs, there are nevertheless many
potential astrophysical sources. Accreting neutron stars
often exhibit X-ray luminosities of $10^{38}$ ergs s$^{-1}$ and
the Crab Pulsar releases its rotational kinetic energy at
a rate of $4\times 10^{38}$ ergs s$^{-1}$. Either could
regularly emit a low-powered FRB without violating
energy conservation laws.

Causality requires that the 
dimension ($d$) of an FRB source must be $d\leq c \delta t$,
where $c$ is the speed of light and $\delta t$
is the duration of the FRB. For observed FRB timescales
less than 1 millisecond, the dimension of the emission region had
to be less than 300 km. This suggested
compact objects such as neutron stars or black holes,
or possibly relativistic shock waves in close proximity of the source at similar scales.
Hot extended plasmas emit incoherent radio emission, due
to the interaction of charged particles, with a spectrum and luminosity determined by their dimension and temperature.
To reach the luminosities of FRBs in the available timescale would require an incoherent source to have 
unphysical temperature ($\sim$ 10$^{40}$ K).
Therefore the FRB emission mechanism must be a coherent process, one where $N$ charged particles emit radio waves all in phase, producing $N^2$ times the power of a single particle\cite{2017RvMPP...1....5M}.
Examples of coherent processes are i) plasma emission involving the
generation of Langmuir waves and subsequent conversion into
waves at the plasma frequency, ii)
electron cyclotron maser emission and iii) pulsars. Of these, the pulsar
emission mechanism is the least understood.
Giant pulses from the Crab Pulsar are thought to be caused by
many nanosecond-timescale shots, each of which individually produces 
coherent emission that appear in quick succession, resulting in a giant pulse\cite{2016MNRAS.457..232C}.

Early models for extragalactic FRBs could be assigned into
two broad categories. In the first, some catastrophic
explosion or other source-destroying event occurred, 
releasing a large amount of energy,
some small fraction of which was converted into a
coherent radio pulse. These are the `cataclysmic' models.
Examples include a neutron star-neutron star merger\cite{2013PASJ...65L..12T},
a core-collapse gamma-ray burst or otherwise unusual (superluminous?) supernova explosion, 
or a neutron star that briefly exceeds its maximum
stable mass before collapsing to a black hole\cite{2014A&A...562A.137F}. 
In these models
the FRB can never repeat, and many of the events 
are expected to be associated with
 star-forming galaxies with large numbers of 
 massive stars and high supernova rates.
Cataclysmic models invoking decelerating blast waves\cite{2019MNRAS.485.4091M} 
predicted unresolved FRBs with radio bandwidths $\delta \nu/\nu\sim 1$ similar
to that of the receivers in many radio telescopes,
whereas models invoking the magnetospheres of relativistic
objects could contain finer temporal and band-limited spectral features\cite{2016MNRAS.457..232C}, as exhibited in 
the giant pulses from energetic pulsars.

The second class of model was non-cataclysmic, so could  allow FRBs to repeat. The term
`giant pulses' refers to the very bright ($\gtrsim$100 $\times$ mean flux density) single pulses emitted by high-magnetic field young 
pulsars including the Crab Pulsar\cite{2003Natur.422..141H}
and PSR J0540--6919\cite{2021MNRAS.505.4468G} that rotate at approximately
20-30 Hz. 
Millisecond pulsars rotate at up to 700 Hz, and 
two examples that emit giant pulses are
PSR B1937+21\cite{1996ApJ...457L..81C}
and the globular cluster pulsar PSR J1823--3021A\cite{2020MNRAS.498..875A}.
FRBs are more than 1 billion times the luminosities of the
most luminous giant pulses from the Crab Pulsar, though it has been
suggested that FRBs could be a related phenomenon producing 
much rarer  
`super giant pulses'\cite{2016MNRAS.457..232C} from energetic pulsars. 
The origin
of the giant pulses is unclear, but if
a pulsar's propensity to emit a giant pulse
depends upon its magnetic field
and spin period, 
there could be extremely magnetic and rapidly
spinning neutron stars in
the Universe (termed millisecond magnetars)
that could emit numerous FRB-like pulses, albeit for
a very short time ($<$ yr) before they exhaust
their rotational kinetic energy.
If this model is correct, FRBs would not be one-off sources,
and (like young pulsars), they would preferentially be located in the spiral arms of 
star-forming galaxies, possibly inside supernova
remnants. They might also
exhibit an underlying quasi-periodicity, like giant pulses which tend to appear at particular rotation phases.
Magnetars (neutron stars with high magnetic fields) would be expected in star-forming
regions but millisecond pulsars are known to occur 
both within globular clusters and the disks and halos of
galaxies\cite{2018MNRAS.481.3966B}, because their rotational kinetic energy is sufficient to power them for more than the age of
the Universe.

Some models did not fit into either category, such as those that implied that FRB sources are within the Milky Way and that the $DM$ was spurious, such as 
 flare stare models\cite{2014MNRAS.439L..46L,2014MNRAS.443L..11D}.
Although these models greatly reduced the required intrinsic luminosities, they 
required an alternative explanation
for the dispersion of the pulses, 
necessitating a fine-tuned 
physical model to produce the observed dispersion sweep. 

\subsection*{Discovery of repeating FRBs}

Searches for repetition of the FRBs detected using Parkes found none, in
 over 100\,h of observation\cite{2015MNRAS.454..457P,2015ApJ...799L...5R}. However in 2016 
 FRB 121102, then the only FRB detected using Arecibo, was
found to repeat, producing many repeat bursts in a single observing session\cite{2016Natur.531..202S}. 
It was nicknamed `the repeater' (later R1). 

In follow-up observations of FRB 121102, ten additional bursts
were discovered, two on one
day, then 8 on another, with six appearing during a ten minute FRB
`storm'. On other days no bursts were seen\cite{2016Natur.531..202S}.
The repeater's discovery ruled out the  cataclysmic
models, at least for for repeating FRBs. The bursts from FRB 121102 also appeared to come in clumps, unlike giant pulses from pulsars that are more random\cite{1995ApJ...453..433L,2020MNRAS.498..875A}.

FRBs from the repeater 
 appeared to be subtly different than the non-repeating FRBs observed with 
Parkes and the GBT. Bursts from the repeater often possessed
multiple components and were broader in temporal extent ($\sim$5 ms vs $\sim$1 ms). 
These components were often confined to 
small fractional bandwidths $ \delta \nu/\nu \sim 0.2 $ with emission within each burst drifting to lower radio frequencies (Figure \ref{fig:Montage} B$'$).
This behaviour had been predicted for radio emission from
magnetars\cite{2002ApJ...580L..65L} before any FRB had been discovered.
The emission has been described like a `sad trombone',
with multiple notes each progressively lower in tone\cite{2020MNRAS.498.4936R}.

The repeater's active periods provided an opportunity
for follow-up with interferometers to determine a precise location suitable
for the identification of host galaxies.
During one such active period,
the realfast\cite{2018ApJS..236....8L} instrument on the Very Large Array (VLA)
interferometer detected an FRB\cite{2017Natur.541...58C}, and 
localised it to near a persistent radio
source and faint optical companion. 
Subsequent very long baseline interferometry further localised
it to its host galaxy\cite{2017ApJ...834L...7T}  
and that it was situated coincident with the persistent radio source\cite{2017ApJ...834L...8M}. 
The results were intriguing. The host
 was a tiny dwarf galaxy containing
40 million solar masses (M$_\odot$) in stars and gas, with a high
specific star-formation rate of about 0.4 M$_\odot$ yr$^{-1}$, and still undergoing its first wave of star formation. The repeater's host galaxy was very similar to those
of many long-duration gamma-ray bursts, and superluminous
supernovae\cite{2020ApJ...899L...6L}. Both of these transients are associated with
young massive stars of low metallicity.
The redshift ($z=0.193$) \cite{2017ApJ...834L...7T} and hence distance to the repeater's host
galaxy removed all remaining doubt
that FRBs were at cosmological
distances.  FRB121102 was located at a distance of
972 Mpc, close to the
maximum estimated for the Lorimer burst
\cite{2007Sci...318..777L}.

The association with
the persistent radio source raised several
questions.
Was the persistent source enabling the FRB emission, caused by the FRB, or merely coincident with it? Perhaps the persistent radio source 
was an accreting intermediate mass black
hole, or nearby supernova remnants powered by young neutron stars produced by recently-exploded stars?

Follow-up observations of the repeater detected 93 further bursts\cite{2018ApJ...863....2G,2018ApJ...866..149Z} between 4 and 8 GHz.
These observations showed that in its rest frame
FRB emission could extend to almost 10 GHz, but often with small fractional
bandwidths. The repeater has a rotation measure of  $\sim$10$^5$ rad m$^{-2}$, among the 
highest known for any astronomical source \cite{2018Natur.553..182M}.
This implied that it was in a highly magnetised environment,
similar to the Galactic Centre of the Milky Way and its supermassive black hole.

Subsequent  observations using broadband receivers \cite{2019ApJ...877L..19G} demonstrated 
that the repeater's emission is often 
limited in frequency extent, with
emission confined to
the same finite radio bands for 
extended periods and must be instrinsic to the source. Monitoring of the
rotation measure has shown it varies over years, dropping to 
2/3 of its original value, indicating a rapidly evolving magnetic environment\cite{2021ApJ...908L..10H}.

Questions posed by the discovery of the repeater
were: Do all FRBs repeat if observed for long enough? Are 
 there two classes of FRB - repeaters and non-repeaters?
 If every FRB source emits
millions of FRBs (or more), the formation rate of the
sources could be very low, potentially highly exotic and completely unknown from observations of our local Universe.

\subsection*{2017-2022: An FRB age of discovery}

The vast majority of the early FRBs were
found with standard radio pulsar search instrumentation. Once
the cosmological population was established plans were made to accelerate the discovery rate, using large field
of view instruments. Some FRBs, such as the Lorimer Burst, were so bright that they should have been detectable by small dishes.
Although small telescopes have much lower sensitivity, they do
have wider fields of view than large dishes and are
less expensive to build and operate. Purpose-built FRB
facilities, both large and small, were
constructed. 

\begin{figure}
	\centering
	\includegraphics[width=1.0\textwidth]{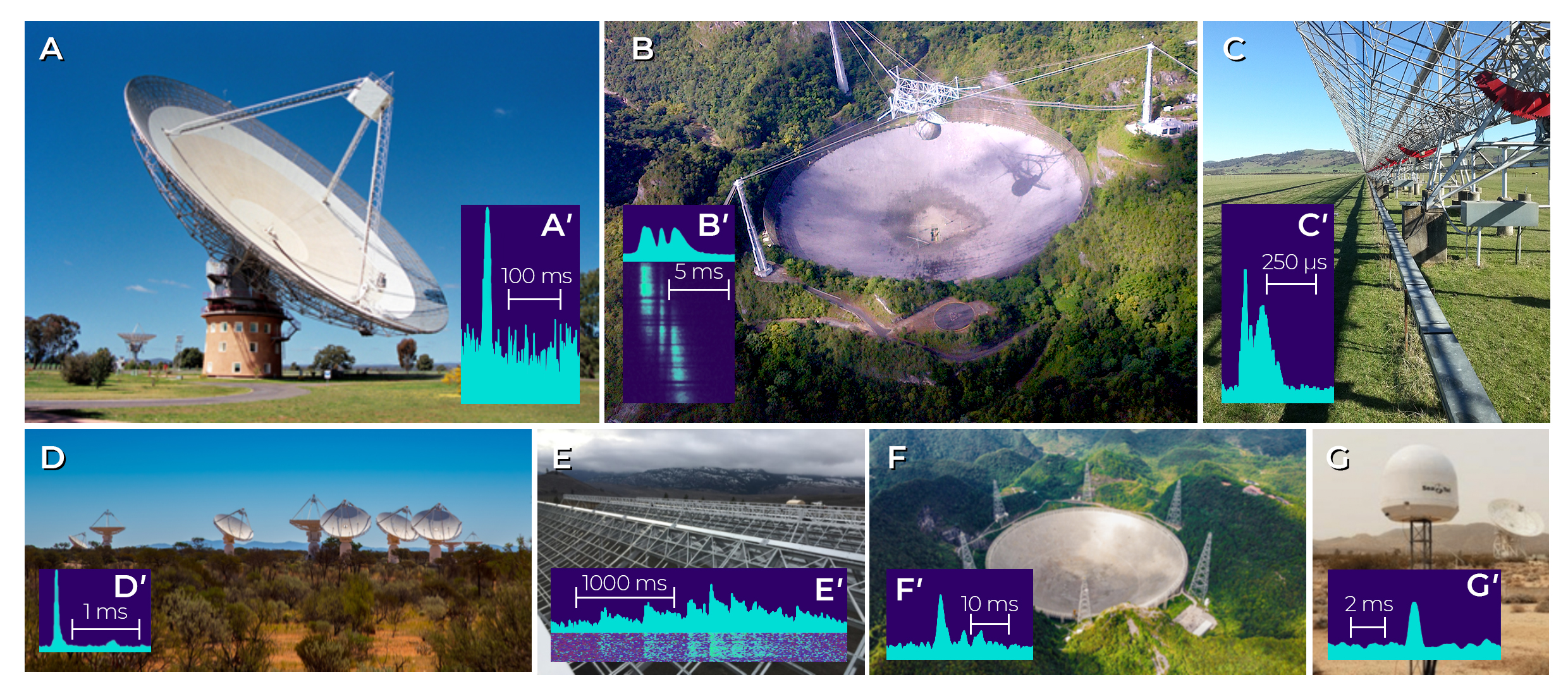} 

	\caption{\textbf{Seven radio telescopes and example FRBs detected at each.} FRBs with interesting frequency structure from Arecibo and CHIME have waterfall plots beneath their integrated profiles.
	    (A) The Parkes 64-m telescope and the Lorimer Burst FRB~010724\cite{2007Sci...318..777L} (inset).
	    (B) The 305\,m Arecibo Observatory and one burst from
	    the repeating FRB~121102\cite{2019ApJ...877L..19G}. The burst  exhibits a downward-drifting frequency effect as a function of time. 
     Frequencies run from low to high in the vertical direction.
	    (C) The 1.5-km long Molonglo telescope
	    and FRB~170827 with its extremely narrow temporal structure obtained by the UTMOST real time data capture system\cite{2018MNRAS.478.1209F,2017PASA...34...45B}.
	    (D) The core of the 36-antenna Australian Square Kilometre Array Pathfinder telescope (ASKAP). The inset shows a four-component  FRB~181112 with narrow temporal features\cite{2020ApJ...891L..38C}.
	    (E) The cylindrical $4\times 20$\,m\,$ \times \,100$\,m Canadian Hydrogen Intensity Mapping Experiment telescope (CHIME) and a 
	    3-second long FRB~20191221A that exhibited a 216.8 ms periodicity\cite{2022Natur.607..256C}.
	    (F) The 500\,m Five-Hundred Metre Aperture Spherical Telescope (FAST) and the three-component fast radio burst FRB~181123 \cite{2020ApJ...895L...6Z}.
	    (G) The Survey for transient radio emission 2 experiment's STARE2 telescope and the Galactic fast radio burst 
	    FRB~200428 it detected from a magnetar during an x-ray flare\cite{2020Natur.587...59B}. Photo credits: 
J. Sarkissian (A), F. Camilo (B), C. Flynn (C), K. Steele (D), M. Bailes (E), Di Li (F), S. R. Kulkarni (G).
	    }
	\label{fig:Montage} 
\end{figure}

The UTMOST upgrade of the large cylindrical Molonglo telescope (Figure \ref{fig:Montage} C) allowed an interferometer to find the first FRBs
in a blind survey\cite{2017MNRAS.468.3746C}. The
 newly commissioned ASKAP
array  (Figure \ref{fig:Montage} D) added an incoherent fast sampling mode
and with its 30 square degree field of view provided by its phased array feeds began detecting FRBs routinely, finding 20 in a fly's eye survey\cite{2018Natur.562..386S}. 
The phased array feeds 
removed the degeneracy between
FRB flux and (usually unknown) position in the primary beam, enabling measurements of absolute flux densities for one-off FRBs. The nearby high-flux FRBs detected with ASKAP were shown to be local versions of fainter FRBs detected by the more sensitive Parkes
dish in the more distant Universe\cite{2018Natur.562..386S}. 
The expected d log $N$/d log $S$ = --3/2 flux density distribution
was beginning to be consistent with a cosmological population.

Many of these purpose-built facilities also had the ability to 
buffer and store the raw data when an FRB occurred, determined in real time. This enabled microsecond time resolution 
studies of FRB profiles, which revealed detailed microstructure down to only a few tens of microseconds\cite{2018MNRAS.478.1209F,
2020ApJ...891L..38C} (Figures \ref{fig:Montage} C$'$ and D$'$). 
These short timescales indicate
the emission arises from sub-km scales, 
 consistent with gaps in neutron star magnetospheres\cite{1986ApJ...300..500C}.

The CHIME telescope in Canada (Figure \ref{fig:Montage} E) has a very wide field of view
($\sim 200$ sq deg)
and high instantaneous sensitivity (8000 m$^2$ of collecting area). This is complemented by the ASKAP interferometer's sub arc-second localising capabilities  
and the even higher sensitivity provided by the 500\,m FAST dish  (Figure \ref{fig:Montage} F). 
A very different approach was taken by the three
20cm STARE2 coaxial feeds (Figure \ref{fig:Montage} G), 
which have less than 
a millionth of FAST's collecting area, but view the entire sky.

CHIME \cite{2018ApJ...863...48C} is 
a fixed 4 $\times\, $100\,m (long) $\times$ 20\,m (wide) cylindrical interferometer
that forms 1024 coherent beams over the sky, covering almost 200 square
degrees. Its FRB sub-project CHIME/FRB searches for dispersed radio pulses almost continuously from 400-800 MHz. It scans the entire northern
sky every day searching for FRBs.
The CHIME/FRB average detection rate 
(a few per day) has rapidly increased the catalogue of known FRBs which now has over 600 unique sources. CHIME observations have provided insights into FRB emission at low (400-800 MHz) radio frequencies\cite{2019Natur.566..230C} and identified a large number of repeaters\cite{2019Natur.566..235C,
2019ApJ...885L..24C,
2020ApJ...891L...6F}, many of which were localised with other facilities.

In early 2020 CHIME detected two FRB-like bursts of 
emission\cite{2020Natur.587...54C} from the Galactic magnetar SGR 1935+2154 
separated by only 30 ms. Just a second earlier, the STARE2
dipoles also detected a ms-duration radio burst\cite{2020Natur.587...59B} (Figure \ref{fig:Montage} G$'$)
at 1.4 GHz. The time delays between the two instruments were consistent with the delay due to pulse dispersion.
These radio bursts were coincident with an X-ray burst from the magnetar\cite{2020GCN.27665....1S}.
Although the radio luminosity of the burst was 30 times
weaker than the least luminous FRB then known, it demonstrated
that magnetars could emit FRB-like emission. At least some
FRBs, and possibly all, are emitted from magnetars.

The once trillion-fold luminosity gap between the Lorimer Burst and the Galactic pulsars 
has gradually closed, though a gap still persists
(Figure \ref{fig:Phase}).
The definition of what constitutes an FRB is also becoming blurred. Early observations of FRBs were often 
temporally smeared by the instrumentation and had $\sim$ms durations but, as instrumentation improved,
both intrinsically narrower and broader FRBs have been
observed. Periodic emission\cite{2022Natur.607..256C}
has also been reported in an unusually long $\sim$3\,s burst of radio emission, with periodic spikes
at separations of 216.8 ms (Figure \ref{fig:Montage} E$'$). 
Two other
FRBs have potential periodic emission with periods of 2.8 and 10.7 ms\cite{2022Natur.607..256C}. Do these reflect the rotation periods of neutron star hosts, or is FRB like that of some radio magnetars, that often exhibit spiky emission\cite{2012MNRAS.422.2489L}?

\subsection*{Repeaters galore}

CHIME detected a 
second repeater\cite{2019Natur.566..235C}  
(FRB 20180814A) with a low $DM$=189 pc cm$^{-3}$ indicating a distance of only 350 Mpc. The detections increased so rapidly
that there were soon
another 17 repeaters\cite{2019ApJ...885L..24C,2020ApJ...891L...6F}. In 
its first year of operation, CHIME detected a total of
18 repeaters (3.7\% of
the total) and 474 FRBs
that were not seen to repeat\cite{2021ApJS..257...59C}.
The repeating FRBs have narrower
fractional radio bandwidths and are wider than one-off 
FRBs\cite{2021ApJ...923....1P}.
Follow-up  
interferometric observations\cite{2022Natur.602..585K} demonstrated that one of these repeaters 
(FRB 20180916B) was in a nearby spiral galaxy at
a redshift of $z=0.0337$ ($\sim$170 Mpc). This FRB was six times closer than the original repeater (R1) and less
luminous that previously observed FRBs beyond our galaxy.

\begin{figure}
	\centering
	\includegraphics[width=0.6\textwidth]{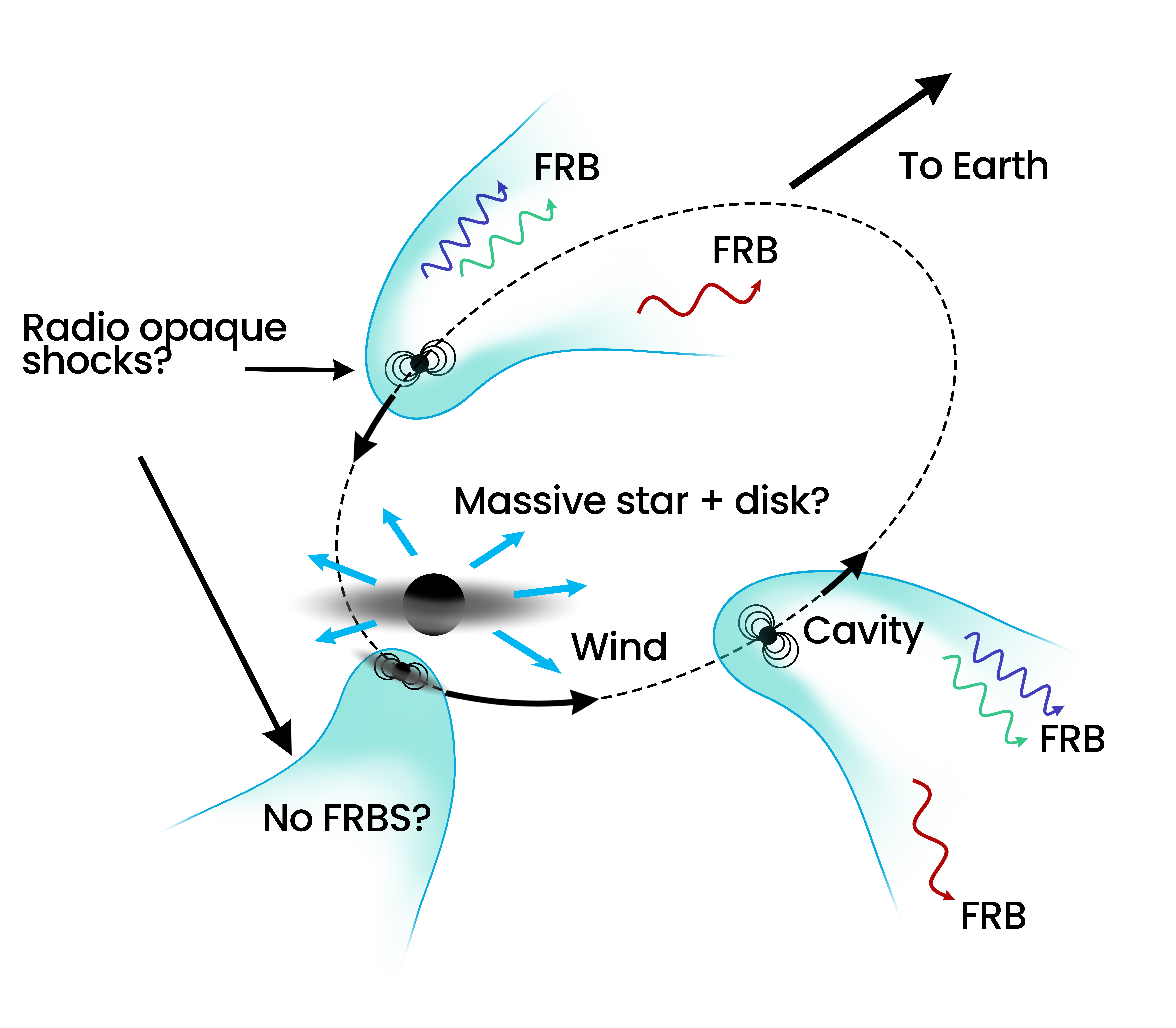} 

	\caption{\textbf{Repeating FRB orbital model.}
		In this repeating FRB model\cite{2017ApJ...836L..32Z}
		a massive star's stellar wind (blue arrows) causes an orbiting magnetar
		(with a period of weeks to months) 
		to emit FRBs. The interaction between the stellar wind and the magnetar wind produces 
		cavities (cyan shading) separated by a shock front
        (cyan lines). The cavity 
  preferentially emits high frequency FRBs (blue and green wavy arrows) 
		at the leading edge, and lower frequency FRBs (red wavy arrows) in the trailing
		sections. This model is an attempt to explain both the activity windows of some
		repeaters and their radio frequency time dependence.
		}
	\label{fig:OB} 
\end{figure}

Analysis of the repeater FRB 180916.J0158+65 
showed the bursts were all received in a 
5\,day-wide window which recurred every 16.35 days, with over half concentrated in a narrower 0.6-day wide window\cite{2020Natur.582..351C}. This motivated searches for periodicities in other repeaters.
R1's bursts 
were then shown to be consistent with a 157-161\,d periodicity, with
a broader fractional activity cycle ($\sim$50\%)
\cite{2020MNRAS.495.3551R,2021MNRAS.500..448C}.
A consistent periodicity usually indicates either i)
 two stars in an orbit, or ii) precession, the
reorientation of a spin axis (tracing out a conical shape). 
This led to the suggestion that repeating FRBs were
magnetars orbiting other active (massive?) stars, in such
a way so that they are only observable
at certain orbital phases\cite{2020ApJ...893L..39L}. This model received some support when
the radio frequency of bursts from repeating FRBs was shown to be (orbital?) phase dependent,
with the lower frequency FRBs coming later in the cycle than the high-frequency
ones \cite{2021ApJ...911L...3P,2021Natur.596..505P}. Others\cite{2017ApJ...836L..32Z} have
suggested that repeating FRBs might be induced by material streaming past magnetars.
These types of models are illustrated in Figure \ref{fig:OB}. 

Recent observations with the Five-hundred metre Aperture Telescope (FAST) in China 
( Figure \ref{fig:Montage} F) detected 1652 FRBs
from the original repeater in less than 60\,h of observation time\cite{2021Natur.598..267L}.
During its most active observed period,
the repeater was bursting on average every 30 seconds.
No periodicity was found, 
and the high repetition rate
means that the radio emission mechanism must be very efficient.
The $DM$ of the repeater is increasing, not decreasing as would be
expected if it was at the centre of an expanding
supernova remnant\cite{2018ApJ...868L...4M,2018ApJ...861..150P}.

Another repeater was found with a $DM$ of only 87 pc cm$^{-3}$,
associated with the spiral galaxy Messier 81 (M81)  only $\sim$3.6 Mpc distant\cite{2022Natur.602..585K}. 
Interferometry helped determine that it is almost certainly 
associated with a globular cluster in M81. 
Most of the dispersion for this burst arises from the 
foreground Milky Way, and within M81, not the intergalactic medium. 
This FRB is only about 0.4 percent of the distance to the original
repeater, of similar brightness and hence about 5 orders of magnitude less luminous.
Globular clusters do not contain young stars, and any magnetars
formed in the first wave of star formation are expected to
have become inactive long ago.
If this repeating FRB is also produced by a magnetar, it might
have been formed recently, either by the 
formation of a neutron star by the collapse of an
accreting white dwarf or
the merger of two neutron stars\cite{2022Natur.602..585K}. The
FRBs might alternatively arise from
a millisecond pulsar, which are abundant in globular clusters
and are known to emit giant pulses\cite{2001ApJ...557L..93R,2020MNRAS.498..875A}.
Follow-up observations of this source found `burst storms',
in which it emits an FRB more than once per minute, but with no associated periodicity\cite{2022NatAs...6..393N}.
Unlike some of the repeaters that show dispersion measure variations, this
source has a stable $DM$ \cite{2022NatAs...6..393N}, consistent
with the expected environment within a globular cluster.

A very high $DM$= 1205 pc cm$^{-3}$
repeating FRB observed with FAST was subsequently localised to a galaxy\cite{2022Natur.606..873N} which is much closer than
the intergalactic $DM$ model suggested, implying that the host galaxy must be contributing $\sim$75\% of the total dispersion. 
Like the original repeater, this FRB is
associated with a persistent radio source - probably
related to the anomalous $DM$. This source
demonstrates the potential pitfalls of assuming the $DM$ provides an accurate cosmological distance.

The linear polarisation fraction of repeating FRBs is strongly radio-frequency dependent as predicted if their radio waves are scattered in a highly variable magnetic environment \cite{2022Sci...375.1266F}. At lower frequencies, radio waves experience more variable Faraday rotation, leading to the
observed systematic depolarisation.

\subsection*{Host Galaxies and Cosmological Applications}

Advances in instrumentation have enabled interferometers to determine the precise locations of
one-off FRBs, identifying their host 
galaxies\cite{
2019Sci...365..565B,
2019Natur.572..352R, 
2020ApJ...899..161L
}, as well as following up repeaters.
A study of 6 repeating and 
10 non-repeating FRBs with known 
host galaxies and redshifts (ranging from
$z=0.008$-$0.66$) found that although there may be hints that
their host galaxies are dissimilar, the differences are
not statistically significant\cite{2022AJ....163...69B}.
FRBs are rare in `red and dead' elliptical galaxies, being
more common in galaxies that are experiencing at least
some star formation\cite{2022AJ....163...69B}. One-off FRBs
are less common in galaxies with high star formation rates per
unit mass, which is unlike the long duration gamma-ray bursts that
are often associated with low-metallicity, low-mass hosts
and produced by exploding massive stars.
Non-repeaters also appear to have different host galaxy
properties to core-collapse supernovae\cite{2022AJ....163...69B}. This
is inconsistent with
unification models that propose all FRBs
are from young magnetars produced in recent supernovae. 
Could non-repeaters be produced by neutron stars re-activated 
long after their formation?
One potential reactivation mechanism is
mass transfer from a companion star. 
This is known to form
the millisecond pulsars, some 
of which emit giant pulses. 
Neutron stars
in binaries accrete mass and gain angular momentum when their
companions exhaust their fuel and swell up 
during stellar evolution.
The length of the delay
between neutron star birth and its `recycling' depends 
upon the mass of the companion star and can be between 1 Myr
and several Gyr. Millisecond pulsars could explain the host properties
of one-off FRBs but
not their lack of repetition.

In cosmology, enough host galaxies have been determined to derive the FRB 
redshift-$DM$ relation\cite{2020Natur.581..391M} (also known as the Macquart relation). 
The observed relation is consistent with the total mass of baryons inferred by studies of the cosmic microwave background \cite{2018ApJ...855..102C,2020A&A...641A...6P}, possibly resolving the difficulty in locating all the baryons using other methods \cite{2012ApJ...759...23S}.

The sheer numbers of FRBs detected by CHIME are providing other
insights into the population.
Analysis of the catalogue\cite{2021ApJS..257...59C} found that
the $DM$s and flux density distributions are consistent
with a cosmological 
population and correlated
with the large scale structure of galaxies 
in the Universe\cite{2021ApJ...922...42R}. 

FRB dispersion measures and host galaxy redshifts have been used to independently derive the value of the Hubble constant ($H_0$). $H_0$ is the rate at which the expansion of the Universe increases with distance.
With a limited sample of 9 FRBs with redshifts, a value of $H_0=$62$\pm$9 km s$^{-1}$ Mpc$^{-1}$ has been deduced\cite{2022MNRAS.511..662H}, albeit with some possibly optimistic assumptions about the (contaminating) host galaxy $DM$ contributions. This is less precise than other methods;
improvements will require eliminating FRBs with high local $DM$ contributions, possibly by examining their intrinsic widths and scattering or by characterising their 
host galaxy environments.

\subsection*{Current status and future prospects}
\begin{figure}
	\centering
	\includegraphics[width=1.0\textwidth]{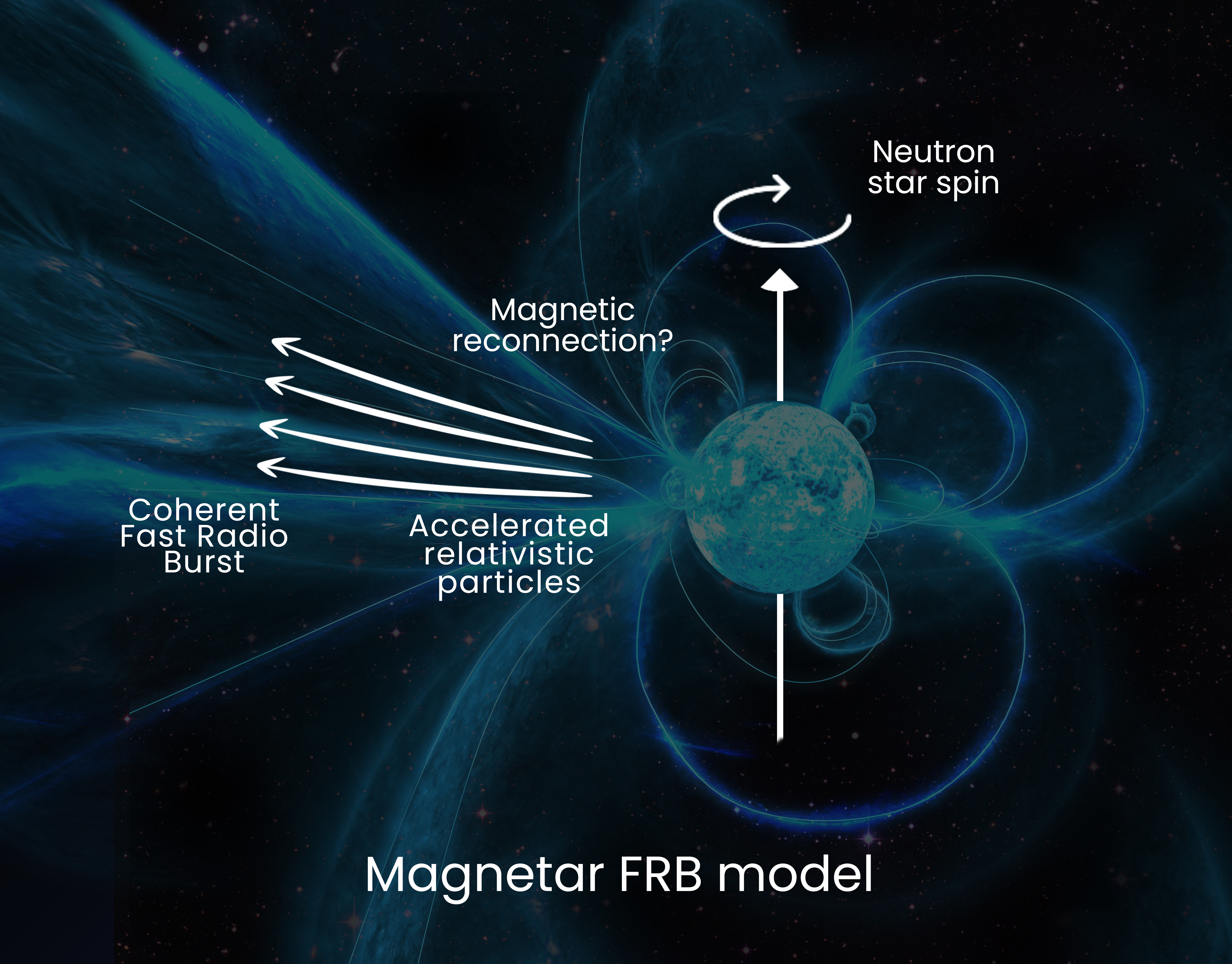} 

	\caption{\textbf{Magnetar FRB emission model.}
		 Reconfiguration of the intense magnetic fields around a magnetar is
		 associated with high-energy outbursts.
        In this model for FRB generation, reconfiguration of the magnetic field
        releases relativistic particles that generate coherent
        radio emission in the magnetosphere, possibly producing fast radio bursts.
		}
	\label{fig:Magnetar} 
\end{figure}

So what are FRBs? 
My personal view is that like many new classes of object, FRBs
will ultimately be shown to be comprised of one or two dominant
sources, but there could be other rarer classes of source with the right combination of magnetic field, rotation, gravity and accelerated charged particles to generate FRBs.
Determining the locations of $\sim$100 FRBs should provide sufficient information on their host
galaxies to constrain the progenitors. My leading contenders
for the repeaters are magnetars, with some in orbits around massive stars, while non-repeaters seem more likely to be  rare giant pulses from 
high-magnetic field ($\sim10^9$ G?) or recently spun-up 
millisecond pulsars. Both magnetars and millisecond pulsars
experience magnetic field reconfigurations (Figure \ref{fig:Magnetar}) leading to changes in 
radio pulse shape changes. In magnetars this can produce 
high-energy
outbursts\cite{2018MNRAS.474..961C} and in one case a low-luminosity FRB\cite{2020Natur.587...59B,2020Natur.587...54C}. Could
these magnetic reconfigurations be a common trigger for FRB production?
In millisecond pulsars these reconfigurations are
extremely rare, and of the few hundred known millisecond pulsars
only a few have exhibited them, including
PSR J1713+0747\cite{2021ATel14642....1X}. 
If this model is correct eventually all FRBs 
might repeat, but we might have to wait decades or more to
observe them.

I expect progress in the field will be strongly
linked to new facilities coming online in the next decade.
The 
MeerTRAP\cite{2016mks..confE..10S}
experiment is expected to detect and localise FRBs at higher
distances. 
CRACO (the CRAFT Coherent upgrade) to the ASKAP interferometer (operational late-2022) will coherently add the signals from the inner 30 antennas to improve the 
FRB localisation
rate by an order of magnitude, 
whilst the Deep Synoptic Array (DSA\,110)\cite{2019MNRAS.489..919K} (late 2022) is predicted to localise
almost one FRB per day. The CHIME/FRB outrigger project 
is deploying additional cylinders to enable localisation. 
CHORD\cite{2019clrp.2020...28V} will be an  array of 512 6-m dishes,
supported by outrigger stations, that will enable rapid localisation
of FRBs and should be operational some time 
in the mid 2020s.
Further in the future instruments like 
the DSA\,2000\cite{2019BAAS...51g.255H} (circa 2027) plan to localise 100s 
of FRBs per day
and from 2030 the Square Kilometre Array\cite{2020MNRAS.497.4107H} is expected to probe high-redshift FRBs.
The results from these new instruments will ensure
the golden age of FRB discovery extends well into the 2030s.

Would FRBs have ever been 
discovered if not for the brightness of the Lorimer Burst? 
Parallels have been drawn with
the first gravitational wave source, which remains the
one with the largest known amplitude\cite{2016PhRvL.116f1102A,2021arXiv211103606T}.
The Lorimer Burst was only detected because
it appeared in the sidelobes of the telescope, so in that sense
its high flux density was necessary for the discovery. But the
high fluence of the burst also compromised both its detection and acceptance. 
The simple interference rejection algorithm in operation tried to erase the burst,
and its brightness led to arguments that it was statistically
unlikely, and hence probably interference.
I'm convinced that the visual recognition of the Lorimer
Burst's dispersed bright pulse played
a crucial role in the belief that FRBs existed,
but I am also certain that it was their 
scientific potential that motivated scientists to
pursue the instrumental developments and surveys that ultimately produced
the scientific discoveries discussed in this review.




\clearpage 
%
\bibliography{science_template} 
\bibliographystyle{sciencemag}

%
%
%
%
%
%


\section*{Acknowledgments}
I thank Duncan Lorimer for involving me in the pursuit of the Lorimer Burst
and to my colleagues in the HTRU, SUPERB, UTMOST and CRAFT FRB collaborations.
I'm grateful to Manisha Caleb and Carl Knox for help with the
figures and Adam Deller, Kelly Gourdji, Chris Flynn 
and Ryan Shannon for feedback on the manuscript. Scott Ransom and another anonymous referee provided many important suggestions that greatly improved the manuscript.
 
\paragraph*{Funding:}
I am grateful to the Australian Research Council for support 
via Laureate Fellowship FL150100148 and the ARC Centre of
Excellence for Gravitational Wave Discovery (OzGrav) grant CE170100004. 
\paragraph*{Competing interests:}
I declare no competing interests.


\end{document}